\documentclass[reprint, aps, prb, floatfix]{revtex4-2}

\usepackage[dvips]{graphicx}
\usepackage[english]{babel}
\usepackage{amsmath}
\usepackage{amssymb}
\usepackage{bm}
\usepackage{color}
\usepackage{subfigure}
\usepackage[dvipsnames]{xcolor}
\usepackage[colorlinks, citecolor = black, linkcolor = black, urlcolor = NavyBlue]{hyperref}

\newcommand{\bs}{\boldsymbol}

\begin{document}
\title{ Interference of Holon Strings in 2D Hubbard Model}
\author{Chang-Yan Wang}
\affiliation{Department of Physics, The Ohio State University, Columbus, OH 43210
}
\author{Tin-Lun Ho}
\affiliation{Department of Physics, The Ohio State University, Columbus, OH 43210
}
\begin{abstract}
The 2D Hubbard model with large repulsion is a central and yet unsolved problem in condensed matter physics for decades. The challenge appears below half filling, where the system is a doped antiferromagnet. 
In this regime, the fermion excitations are nothing like those in a Fermi liquid, which carry both spin and charge. Rather, they split up into holons and spinons, carrying charge and spin separately. Moreover, the motion of a holon is believed to stir up the underlying antiferromagnetic order, leaving behind it a string of ``wrong" spins. While direct observation of the holon string is difficult in electron systems, it has become possible in cold atom experiments due to recent experimental advances. Here, we point out the key feature of the holon strings, i.e. its Marshall phase, can be observed through measurements of spin correlations. Moreover, the interference of these strings leads to an anisotropic holon propagation clearly distinguishable than those of spinless fermions, as well as a large suppression of the magnetic order in the region swept through by the strings, as if the system is driven towards a spin liquid. 
We further illustrate the effect of the Marshall phase by showing the motion of a holon in the so-called $\sigma tJ$-model where the Marshall phase is removed. 
\end{abstract}

\maketitle

\section{Introduction}
The Fermi Hubbard model is one of the most important model in condensed matter physics. It has been intensely studied since its appearance\cite{hubbard_electron_1963}. The model has a very simple form, consisting  a term describing electron hopping on a lattice and a term describing short range interaction between electrons. There is considerable evidence showing  that the model can exhibit a wide range of phenomena in various parameter regime, such as  metal-insulator transition, antiferromagnetism, superconductivity, etc \cite{tasaki_hubbard_1998}. The 2D Hubbard model with strong repulsion is of particular interest, as it is believed to capture the key physics of high-$T_c$ superconductivity \cite{anderson_theory_1997}. Yet despite decades of studies, the model remains unsolved. 

For simplicity, we shall consider the 2D Hubbard model on a square lattice. For strong repulsion, the system at half filling has one electron per site, with an  antiferromagnetic (AF) ground state. The central question has been  how this ground state is changed as the system is doped below half filling. The problem is challenging because the fermion excitations of a doped AF are believed to split up into "holons" and "spinons", which carry charge and spin separately. They are 
very different from the excitations of a Fermi liquid, which carry both spin and charge. 
The unusual properties of the spinons and holons are believed to be the cause of many unusual properties of high $T_c$ superconductors including the ``strange metal" behavior\cite{anderson_theory_1997}. Yet despite their central role in theoretical studies, there are no known ways to observe them {\em directly} in current solid state experiments. However, an exciting possibility has emerged in the last few years. With the rapid advances in cold atom experiments, one can now simulate with great precision the Fermi Hubbard model using ultra-cold fermions in optical lattices, and observe the AF order at half filling\cite{mazurenko_cold-atom_2017,hilker_revealing_2017,greif_short-range_2013,cheuk_observation_2016,hart_observation_2015}. 
Moreover, with the development of atom microscope that  can image atoms with single site resolution, one can now image the many-body wavefunction of a quantum state with unprecedented detail. 

In the case of a holon, it is supposed to leave behind it a string of ``wrong" spins as it moves through the AF background.  Recently, Markus Greiner's group tried to identify the holon strings by comparing the images of a doped AF with a classical AF background, and had concluded their presence by comparing the observed images with a theoretical model \cite{chiu_string_2019}.
Propagation of holon has also been studied in \cite{bohrdt_dynamical_2020, grusdt_parton_2018, grusdt_microscopic_2019}.
More recently, one of us (TLH) has introduced a method to identify the holon string directly from the spin density of the system without making reference to the results of specific theories\cite{ho_imaging_2020}. The method  relies only on an exact property of the Heisenberg AF -- that its ground state obeys the Marshall sign rule.  In ref.\cite{ho_imaging_2020}, it is showed that if the holon moves along the $x$ or $y$ axis of a square lattice, then its string has a very clear signature --  that the spin correlation of neighboring sites is ferromagnetic in the immediate vicinity of the string,   while being AF everywhere else. This behavior follows from a key property of the holon string -- that it carries a phase reflecting the Marshall sign of the fluctuating AF background. We shall refer to this phase as the ``Marshall" phase\cite{sheng_phase_1996, ho_imaging_2020}.  

 In this paper, we focus on the interference effect of holon strings and how they are reflected in the spin correlation of nearest neighbors. The interference effects occur  when a holon travels away from the $x$ or $y$  axis. In this case, there are many strings of the same length connecting the initial and the final position of the holon. As we shall see, the different Marshall phases of different strings will lead to a strong suppression of nearest neighbor spin correlations, as if driving the system towards a spin-liquid. We should mention that the ground state wavefunction of a holon has also been studied for finite systems with periodic boundary condition \cite{chen_single-hole_2019}  and for ladders systems \cite{zhu_spin_2016}. Here, rather than focusing on the ground state, we study the propagation of holons in physical environments created in current experiments. After this study on the interference of strings, we 
further demonstrate the effect of the Marshall phase by comparing our results with those of the so-called $\sigma tJ$-model \cite{zheng_charge-spin_2018}, which is the Hubbard in strong repulsion limit but with the Marshall phase stripped off.  In this case,  a single holon in an AF background behaves more like a fermion in an empty lattice, as if recovering the Fermi liquid behavior.  In other words, the Marhshall phase is responsible for the non-Fermi behavior of the holon.

\section{\texorpdfstring{ $tJ$}{} model, Marshall sign, and amplitudes for holon propagation}
We start with the Hubbard model, 
\begin{eqnarray}
  H = -t \sum_{<i,j>, \sigma} c_\sigma^\dagger(i) c_\sigma(j) + U\sum_i n_\uparrow (i) n_\downarrow (i)
\end{eqnarray}
The first term describes the hopping of a fermion $c_\sigma(j)$ with spin $\sigma$ hopping from site $i$ to a neighboring sites $j$, $t>0$, $U$ is the on-site interaction between opposite spins, and $n_\sigma(i) = c_\sigma^\dagger(i) c_\sigma(i)$. For strong repulsion ($U >> t$), each site can at most be occupied by one fermion. The Hubbard model then reduces to the $tJ$-model $H_{tJ} = \mathcal{T} + H_J$\cite{ }:
\begin{eqnarray}
  H_J = J \sum_{<i,j>} \mathbf{S}_i \cdot \mathbf{S}_j,\,\,\,\, J = t^2/U > 0, \hspace{1.2in} \\
  \mathcal{T} = -t\sum_{<i,j>,\sigma} \overline{c}_\sigma^\dagger (i) \overline{c}_\sigma (j),\ \overline{c}_\sigma (i) = c_\sigma(i)(1 - n_{-\sigma}(i)), \hspace{0.3in}
\end{eqnarray}
where $\mathbf{S}_i = c_\mu^\dagger(i) \bs{\sigma}_{\mu\nu} c_\nu(i)/2$.  $H_J$ is the nearest neighbor AF Heisenberg Hamiltonian with  spin interaction $J$, and $\mathcal{T}$ is the hopping of fermions subject to the constraint of no double occupancy. This constraint  is the origin of the intricate transport  of the system. At half filling,  $\mathcal{T}$ vanishes. 
The $tJ$-model reduces to the AF Heisenberg model $H_J$. Experimentally, an immobile hole at a selected site can be created by piercing through it with a focused blued detuned laser. Note that $\mathcal{T}$  remains zero as long as the hole is immobile. 

We shall denote the spin states of the  Heisenberg system with an immobile hole at ${\bf R}$ as $|\bs{\nu}; {\bf R}\rangle$, where $\bs{\nu}\equiv ( 1 \nu_{1},2 \nu_{2}, .. )$ represents the spin configuration where the fermion at site $i$ has spin  $\nu_{i}$, $\nu_{i}=\uparrow, \downarrow$. Let $|G; {\bf 0}\rangle$ be the  ground state of this Heisenberg AF with an immobile hole at the origin. It has the expansion 
\begin{equation} 
|G, {\bf 0}\rangle = \sum_{\bs{\nu}}
G(\bs{\nu}, {\bf 0}) |\bs{\nu}, {\bf 0}\rangle.
\end{equation}
It is shown in ref.\cite{ho_imaging_2020} that the ground state wavefunction $G(\bs{\nu}, {\bf 0})$ obeys the Marshall sign rule as in the hole free case, i.e. it changes sign when  two opposite spins at nearest neighbor sites 
$\langle i, j \rangle$ are interchanged,
\begin{eqnarray}\label{marshall_sign}
  G(i \uparrow, j \downarrow, ... , {\bf 0}) = -G_{{\bf 0}}(i \downarrow, j \uparrow, ... , {\bf 0}), 
\label{Marshall} \end{eqnarray}
where $(\cdots)$ means the configuration of other spins. Because of this property, it is convenient to use the new spin basis $\overline{|\bs{\nu}, {\bf 0}}\rangle = (-1)^{N_{a}^{\downarrow}}
|\bs{\nu}, {\bf 0}\rangle$,
where $N_{a}^{\downarrow}(\bs{\nu})$ is the number of down-spins  in configuration $\bs{\nu}$ in one of the two sub-lattices of the square lattice (denoted as $a$). (See Figure 1). We can  expand any state $|\Phi, {\bf 0}\rangle$ (including the ground state $|G, {\bf 0}\rangle$) as 
$|\Phi, {\bf 0}\rangle = \sum_{\bs{\nu}}
\overline{\Phi}(\bs{\nu}, {\bf 0}) \overline{|\bs{\nu}, {\bf 0}\rangle}$, where 
$\overline{\Phi}(\bs{\nu},{\bf 0}) = (-1)^{N_{a}^{\downarrow}}\Phi(\bs{\nu},{\bf 0})$.  
If $|\Phi, {\bf 0}\rangle$ satisfies the Marshall sign rule,  Eq.(\ref{Marshall}), then we have  $\overline{\Phi}(\bs{\nu})\geq 0$.

To allow the hole to move, we remove the focused laser at time $\tau=0$. The quantum state at later time is  
\begin{eqnarray}
|\Psi (\tau)\rangle = e^{-i\tau H_{tJ} }|G; {\bf 0}\rangle  
= \sum_{\bf R}\sum_{\bs{\mu}} 
\overline{ \Psi}(\bs{\mu}, {\bf R}; \tau) \overline{|\bs{\mu}, {\bf R}\rangle},  
\hspace{0.3in} \label{evolve1} \\
\overline{ \Psi}(\bs{\mu}, {\bf R}; \tau) = \sum_{\bs{\nu}} \overline{\langle \bs{\mu}; {\bf R}|} e^{-iH_{tJ}\tau }
 \overline{|\bs{\nu}; {\bf 0}\rangle} \,\,
 \overline{G}(\bs{\nu}, {\bf 0}). 
 \hspace{0.4in}
 \label{evolve2}
\end{eqnarray}
Although $\overline{G}(\bs{\nu},{\bf 0})$ is positive,  $\overline{\Psi}(\bs{\mu}, {\bf R};\tau)$ needs not to be because of the holon propagator. 

As discussed in ref.\cite{ho_imaging_2020}, the Marshall sign of a  state $\Phi$ shows up in the "exchange overlap" of opposite spins at neighboring sites $\langle i, j \rangle$, $\rho_{ij} = \langle c_\downarrow^\dagger(i) c_\uparrow^\dagger(j)  c_\downarrow(j) c_\uparrow(i) \rangle_\Phi $, 
\begin{equation}
 \rho_{ij} = \sum_{(...)} \Phi(i\downarrow; j\uparrow; ...)^{\ast} \Phi(i\uparrow; j\downarrow; ...) = \langle S_{i}^{+} S_{j}^{-} \rangle_{\Phi}. 
\label{ex} \end{equation}
The satisfaction of Marshall sign implies $\rho_{ij}<0$. For systems with SU(2) symmetry (such as the $tJ$-model), Eq.(\ref{ex}) can be simplified as $\rho_{ij}= 2\langle S_{i}^{z} S_{j}^{z} \rangle_{\Phi}$.  For systems without $SU(2)$ symmetry, such as the $\sigma tJ$-model we discuss later, the exchange correlation can be obtained using the interference method discussed in ref.\cite{ho_imaging_2020}. To measure the spin-spin correlation of the state when the hole has moved to ${\bf R}$ after time $\tau$, one can first takes many images of the spin density after the hole is released for a time $\tau$, and then post select from these images the subset where the hole has arrived at ${\bf R}$. 

\begin{figure}
    \centering
    \includegraphics[width = 0.4\textwidth]{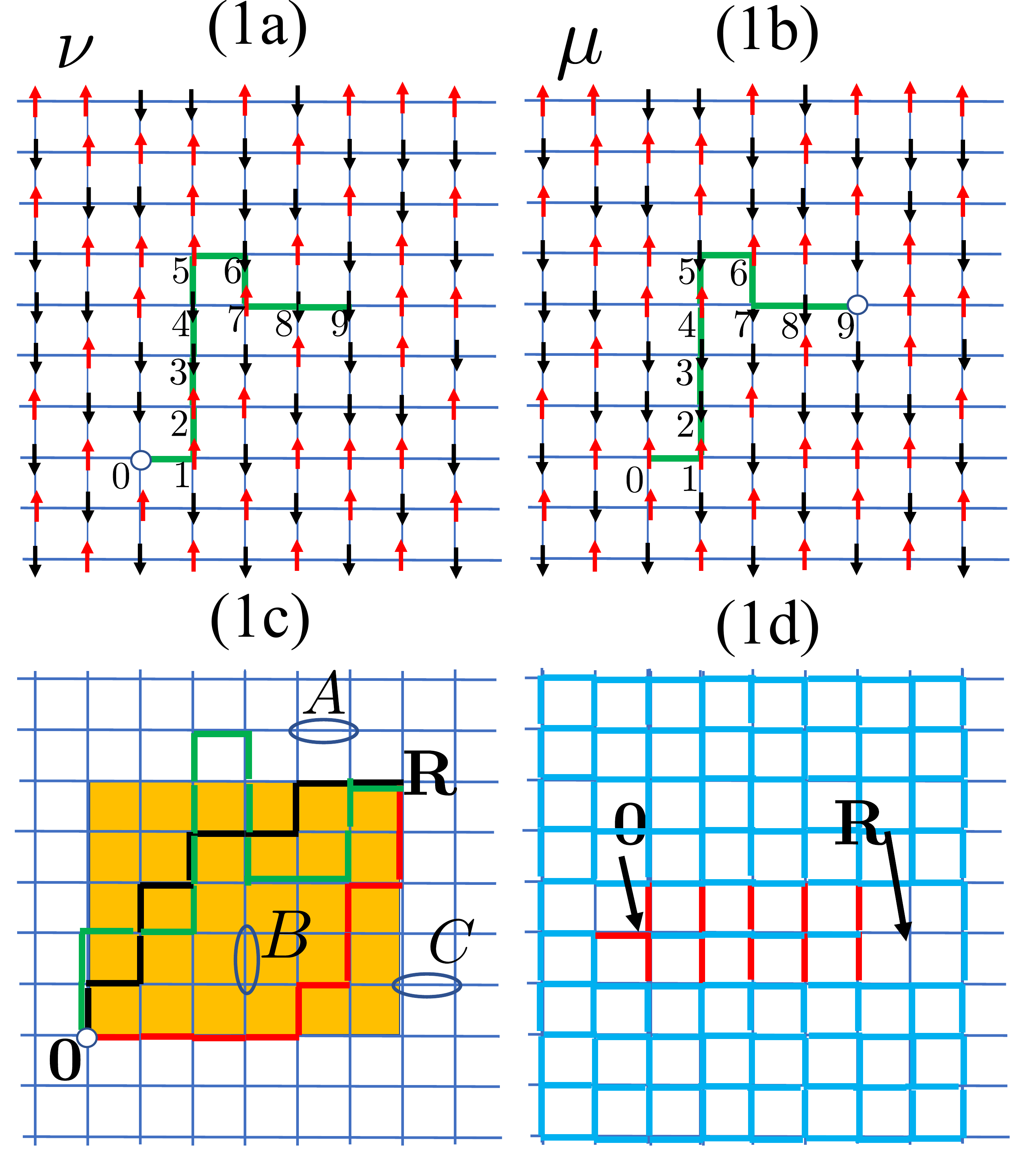}
    \caption{(1a) shows one of the spin configurations $\bs{\nu}$ in the ground state of an antiferromagnet with a hole fixed at site-0.  When the hole is released, it will travel to  sites through various paths. Here, we show a particular path for the holon to travel from site-0 to 9 through the sequence of sites $0,1,2, ..9$, which corresponds to ${\bf R}_{0} ... {\bf R}_{9}$ in Section (III.1), with ${\bf R}_{0}\equiv {\bf 0}$, ${\bf R}_{9}\equiv {\bf R}$. The spin configuration of the initial state and final state are denoted as $\bs{\nu}$  and $\bs{\mu}$. (1b) shows the spin configuration when the holon arrives at ${\bf R}$ through the hopping ${\cal T}$ along the path indicated. This path is one of the terms in Eq.(\ref{Gamma}),  $\prod_{i=1}^{9} \overline{\langle \bs{\mu}_{j}, j|}(-{\cal T}/t) \overline{|\bs{\mu}_{j-1}, j-1\rangle}$. Here, we have $\bs{\mu}_{0}=\bs{\nu}$, $\bs{\mu}_{9} = \bs{\mu}$. From this product, we sees that $\mu_{\ell}=\nu_{\ell}$ for all sites $\ell$ not on the string.  For the sites $i$ on the string, we have $\mu_{i}= \nu_{i+1}$.  The value of the product is $(-1)^{N^{\downarrow}}$ where $N^{\downarrow}$ is the number of down spins on the string \cite{ho_imaging_2020, sheng_phase_1996}. (1c) shows different strings connecting the initial position ${\bf 0}$ and the final position ${\bf R}$ of the holon. Among the three strings (green, black, and red) shown, the latter two have minimum length. The  rectangle $\Lambda$ (orange color) is the area swept through by all the strings of minimum length. Each string has a Marshall phase $(-1)^{N^{\downarrow}}$ defined in  Eq.(\ref{Gamma-sum}). The nearest neighbors (A)  and (B)   have both sites  outside and inside $\Lambda$ respectively. The  The nearest neighbor (C) has one site in $\Lambda$ and one site outside. (1d) shows that the   spin correlation $\langle S^{z}_{i}S^{z}_{j}\rangle$ of the nearest neighbor $\langle i,j\rangle$ after the holon has travelled along a straight line in x direction. All the correlations are antiferromagnetic (blue)  except those in the immediate vicinity of the string, which are ferromagnetic (red). 
    }
\end{figure}

\section{Holon strings in the AF background}

\noindent (III.1) {\em Marshall phase effects on holon propagation: } 
Exact calculation of the propagation amplitude $\Psi(\bs{\mu}, {\bf R};  \tau)$ is formidable. However, the situation is simplified considerably when $J/t \ll 1$, where spins flips are much slower than the motion of holes. In this limit, which is satisfied in current experiments\cite{greif_short-range_2013}, one can expand Eq.(\ref{evolve2}) in powers of $J/t$. To the lowest order, one can replace the $H_{tJ}$ by ${\cal T}$, and Eq.(\ref{evolve2}) becomes 
\begin{equation}
\overline{\Psi}(\bs{\mu}, {\bf R}; \tau) = \sum_{\bs{\nu}}  \Gamma^{}(\bs{\mu}, {\bf R}; \bs{\nu}, {\bf 0}; \tau)\overline{G_{{\bf 0}}}(\bs{\nu})
\label{expand} \end{equation}
\begin{equation}
   \Gamma^{}(\bs{\mu}, {\bf R}; \bs{\nu}, {\bf 0}; \tau) = \sum_{n=0,1,2..} \frac{(it\tau)^n}{n!}  \Gamma^{(n)}(\bs{\mu}, {\bf R}; \bs{\nu}, {\bf 0}; \tau) 
\label{Gamma-sum}\end{equation}
where  $\Gamma^{(n)}$ is the transition amplitude of the hole reaching ${\bf R}$ from ${\bf 0}$ through $n$ nearest neighbor hops, 
\begin{equation}
\Gamma^{(n)}(\bs{\mu}, {\bf R}; \bs{\nu}, {\bf 0}; \tau) 
= \overline{\langle \bs{\mu}, {\bf R}|} \left(\frac{-{\cal T}}{t}\right)^{n} \overline{|\bs{\nu}, {\bf 0}\rangle}. 
    \label{Gamma}    
\end{equation}
The string connecting ${\bf 0}$ to ${\bf R}$ through $n$-hops will be referred to as an $n$-string, and is denoted by the sequence of $n+1$ sites
$({\bf R}_{0},{\bf R}_{1}, .., {\bf R}_{n}) $, where ${\bf R}_{0}={\bf 0}$, ${\bf R}_{n}={\bf R}$. See Figure 1a and 1b.  Successive ${\bf R}$'s are nearest neighbors.
It is clear from Eq.(\ref{Gamma}) that $\Gamma^{(n)}$ is non-vanishing only when the final and initial spin configurations $\bs{\mu}$ and $\bs{\nu}$ are identical ($\bs{\mu}_{i}=\bs{\nu}_{i}$) on all the sites $i$ not on the string. For the spins on the string, the final spin configuration is given by the initial one sliding along the string by one lattice site, i.e. $\mu_{_{{\bf R}_{i}}} = \nu_{_{{\bf R}_{i+1}}}$ for $i=0,1,2, .., n-1$.  
With this relation between the initial and the final spin configurations, the value of $\Gamma^{(n)}$ is \cite{ho_imaging_2020, sheng_phase_1996},
\begin{equation}
    \Gamma^{(n)}(\bs{\mu}, {\bf R}; \bs{\nu}, {\bf 0}; \tau)=
    \sum_{\rm n-string}(-1)^{N^{\downarrow}_{\rm n-string}}, 
\label{Gamma-n-sum}\end{equation}
where $N^{\downarrow}_{\rm n-string}$ is the number of down spins on the $n$-string, and the factor $(-1)^{N^{\downarrow}_{\rm n-string}}$ (which is $\pm 1$) is the Marshall phase of the string. 
\\

From Eq.(\ref{Gamma}), it is clear that $\Gamma^{(n)}$ is non-zero only when the length of the string $n$ exceeds a minimum value $n^{\ast}\geq R_{x}+R_{y}$, which is the minimum number of hops from ${\bf 0}$ to ${\bf R}$ through ${\cal T}$. See Figure 1(c) and 1(d) The number of such strings is $L_{\bf R}=(R_{x}+R_{y})!/(R_{x}!R_{y}!)$. The area swept through by these strings is a rectangle ($\Lambda$) of size $R_x \times R_{y}$. 
If the amplitude $\Gamma$ is approximated by the leading term $\Gamma^{(n^{\ast})}$, we then have 
\begin{equation}
  \overline{\Psi(\bs{\mu}, {\bf R}; \tau)} \propto \sum_{\bs{\nu}}\Gamma^{(n^{\ast})}(\bs{\mu}, {\bf R}; \bs{\nu}, {\bf 0}; \tau) \, \overline{G}(\bs{\nu}, {\bf 0})  
\label{app} \end{equation}
From the constraints on spin configurations imposed by $\Gamma^{(n)}$ we have just discussed, it is clear from Eq.(\ref{app}) that the final state  $\overline{\Psi}$ and the initial state $\overline{G}$ have identical spin configurations for all sites $i$ outside $\Lambda$, i.e. $\mu_{i} = \nu_{i}$. 
 Consequently, the exchange overlap $\rho_{ij}$ (or the spin correlation 
$\langle S^{z}_{i}S^{z}_{j}\rangle$) of both states are the same for all neighboring pairs $<i,j>$ outside $\Lambda$.  (See Figure 1(c)). On the other hand, if $i$ is outside $\Lambda$ and $j$ is inside, then the spin at site $j$  of the two states  $\overline{\Psi}$ and $\overline{G}$ need 
not be the same. This is because some strings in $\Lambda$ will pass through site $j$, each of which will change the spin $\nu_{j}$ of the original AF configuration ($\bs{\nu}$) by the sliding
motion along its   respective path as discussed before, while carrying its Marshall phase. For  strings with opposite Marshall phases,  they interfere destructively and hence weaken the original AF order. 

Similar weakening occurs with both $i$ and $j$ are inside $\Lambda$. However, the case when ${\bf R}$ is on the $x$ or $y$ axis is an exception. In this case, there is only one string of minimum length, which is the straight line connecting ${\bf 0}$ and ${\bf R}$. If both neighboring sites $\langle i, i+1 \rangle$ are on this  line, then the final spin configuration is simply the spin configuration  sliding backward along $x$ by one lattice site., i.e.  $(\mu_{i}, \mu_{i+1}) = 
(\nu_{i+1}, \nu_{i+1})$.  The  AF correlation of neighboring sites are therefore  unchanged\cite{ho_imaging_2020}. (See Figure 1(d)). 

If longer strings  are included in $\Gamma^{(n)}$, (i.e. with $n>n^{\ast}$),  the region $\Lambda$ swept through by the strings will expand beyond the rectangle $R_x \times R_{y}$. 
Still, the region is of finite extent. The behaviors of spin correlations of neighboring sites with different overlap with  $\Lambda$ remain unchanged.  Moreover, since the number of strings grows rapidly as $n$ increases, and their contributions to $\Gamma^{(n)}$ tend to cancel each other. As a result, for time interval $\tau<1/t$, the magnitudes of $\Gamma^{(n)}$ decreases rapidly with $n$. As we shall see, the leading term  $\Gamma^{(n^{\ast})}$ provides a good approximation to the full amplitude $\Gamma$. 

In the next section, we shall present numerical results to demonstrate all the effects mentioned above. We shall show that when the holon propagates away from the $x$-axis, the interference of holon strings strongly reduces the magnitude of spin-spin correlation, as if driving the system towards a spin liquid.   \\

\noindent (III.2) {\em Switching off the Marshall phase, the $\sigma tJ$-model}:   
Another way to demonstrate the Marshall phase is to study the motion of the holon with the Marshall phase is  ``switched  off". 
 This can be achieved by changing the kinetic energy in the $tJ$-model 
to $\widetilde{\cal T}= -t \sum_{<i,j>, \sigma} \sigma \overline{c}^{\dagger}_{\sigma}(i)\overline{c}^{}_{\sigma}(j)$, resulting in the so-called $\sigma tJ$-model \cite{zheng_charge-spin_2018} 
Although we do not yet have a simple way to generate this hamiltonian in cold atom experiments, the model is worth studying because it strips off the Marshall phases of the original model. 
At half filling, $H_{\sigma tJ}$ again reduces to the Heisenberg AF since $\widetilde{\cal T}$ also vanishes. The amplitude for the 
 propagation of a hole is still given by $\Gamma$ and $\Gamma^{(n)}$ in Eq.(\ref{Gamma-sum}) and (\ref{Gamma}) with ${\cal T}$ replaced by $\widetilde{\cal T}$.   
 It is easy to see that when a hole travels from ${\bf 0}$ to ${\bf R}$ through a particular path (or $n$-string), the hopping $\widetilde{\cal T}$ accumulates a phase $(-1)^{N^{\downarrow}_{\rm n-string}}$ cancelling  exactly the same factor in Eq.(\ref{Gamma-sum}) arising from AF background. 
 Consequently, all strings amplitudes add coherently, and the amplitude  $\langle {\bf R}| (-T/t)^{n}|{\bf 0}\rangle$ is simply the total number of strings connecting ${\bf 0}$ and ${\bf R}$. This feature is identical to  that of a spinless fermion in an empty square lattice, although in that case there is no spin sums. 
 Due to the difference in the phase coherence of the strings, the behavior of holons in the $\sigma tJ$ model is very different from that in the $tJ$-model, which we shall show in the next section.    

\begin{figure}
  \subfigure[]{\label{proba_tj}
    \includegraphics[width = 0.35\textwidth]{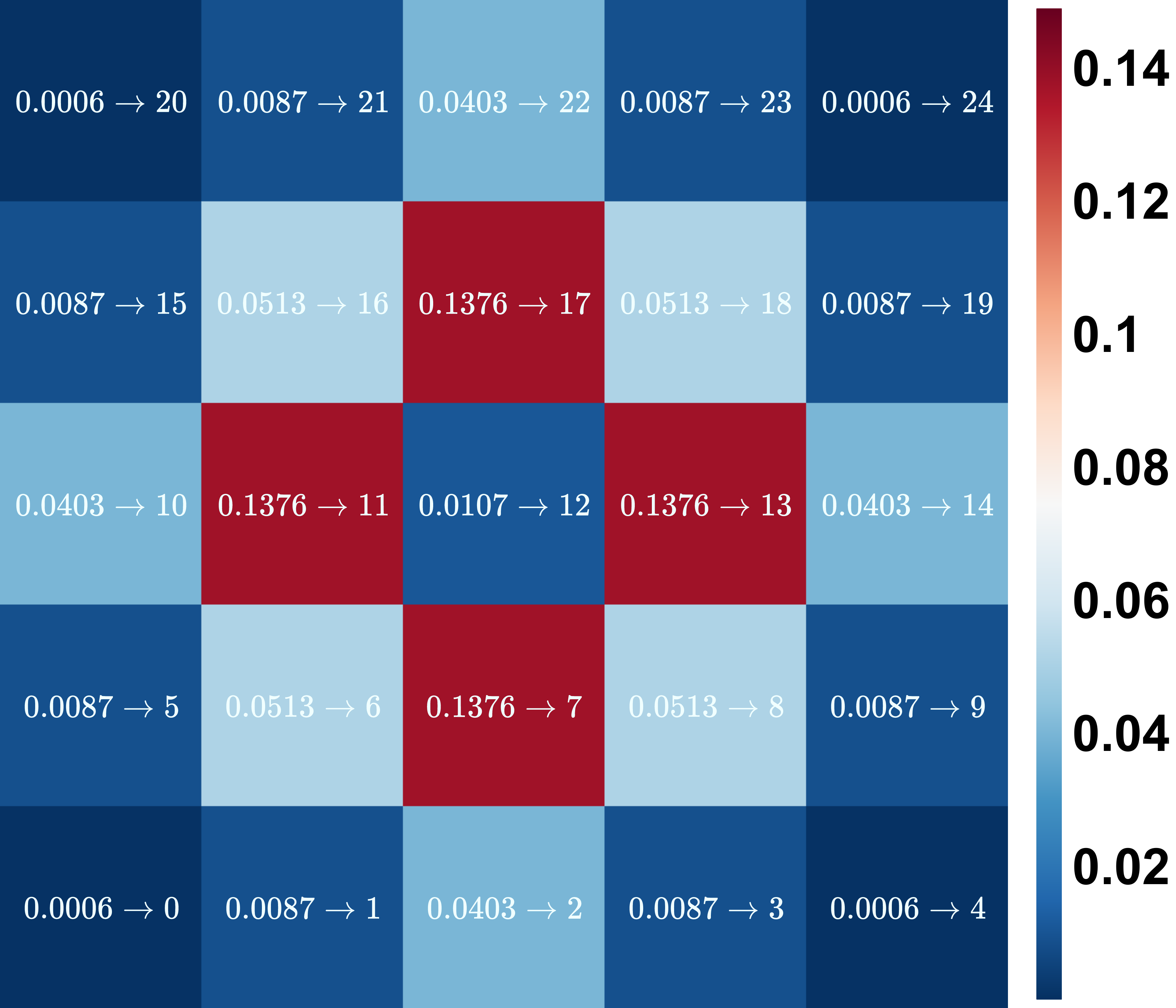}
  }
  \subfigure[]{\label{proba_sig}
    \includegraphics[width = 0.35\textwidth]{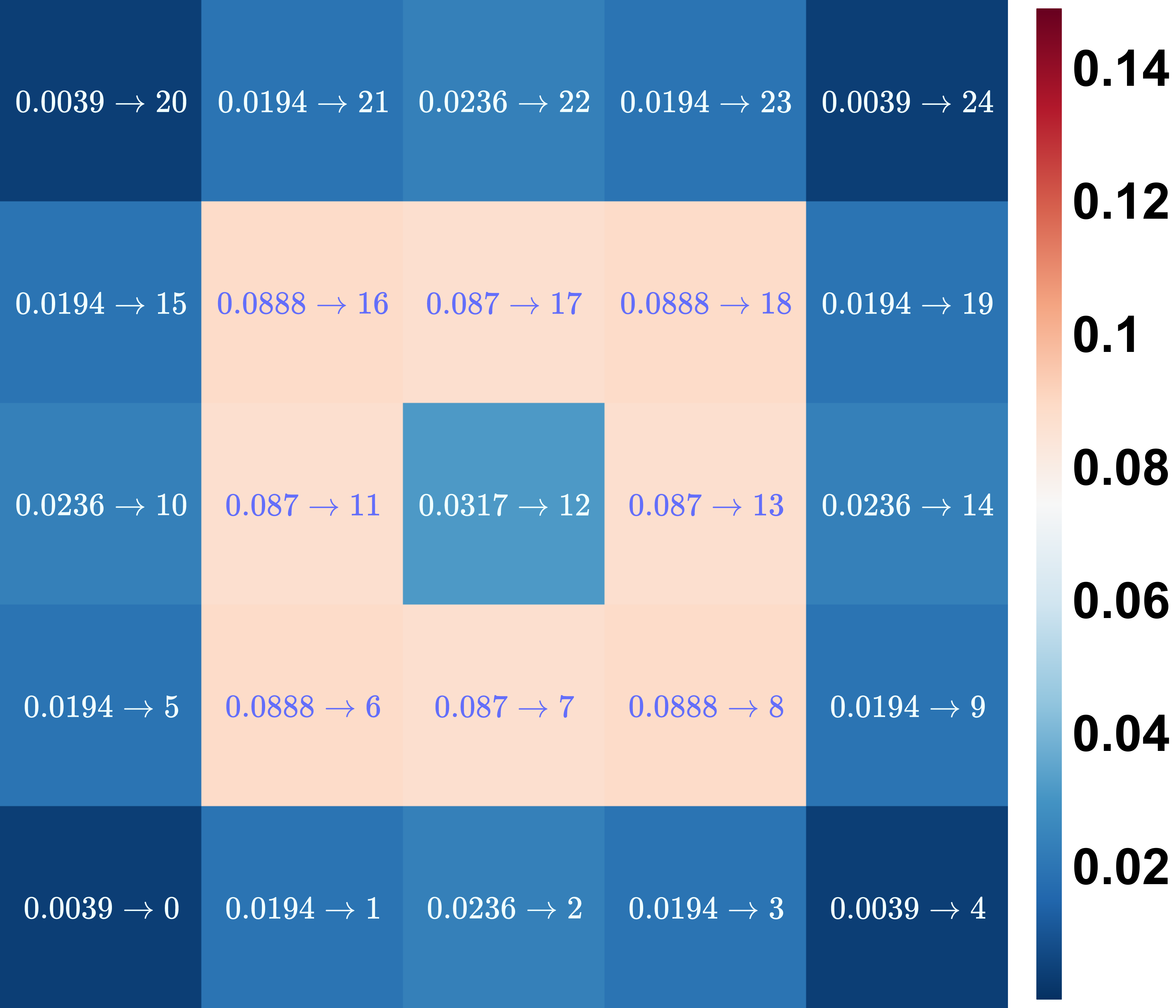}
  }
  \subfigure[]{\label{electron_possib}
    \includegraphics[width = 0.35\textwidth]{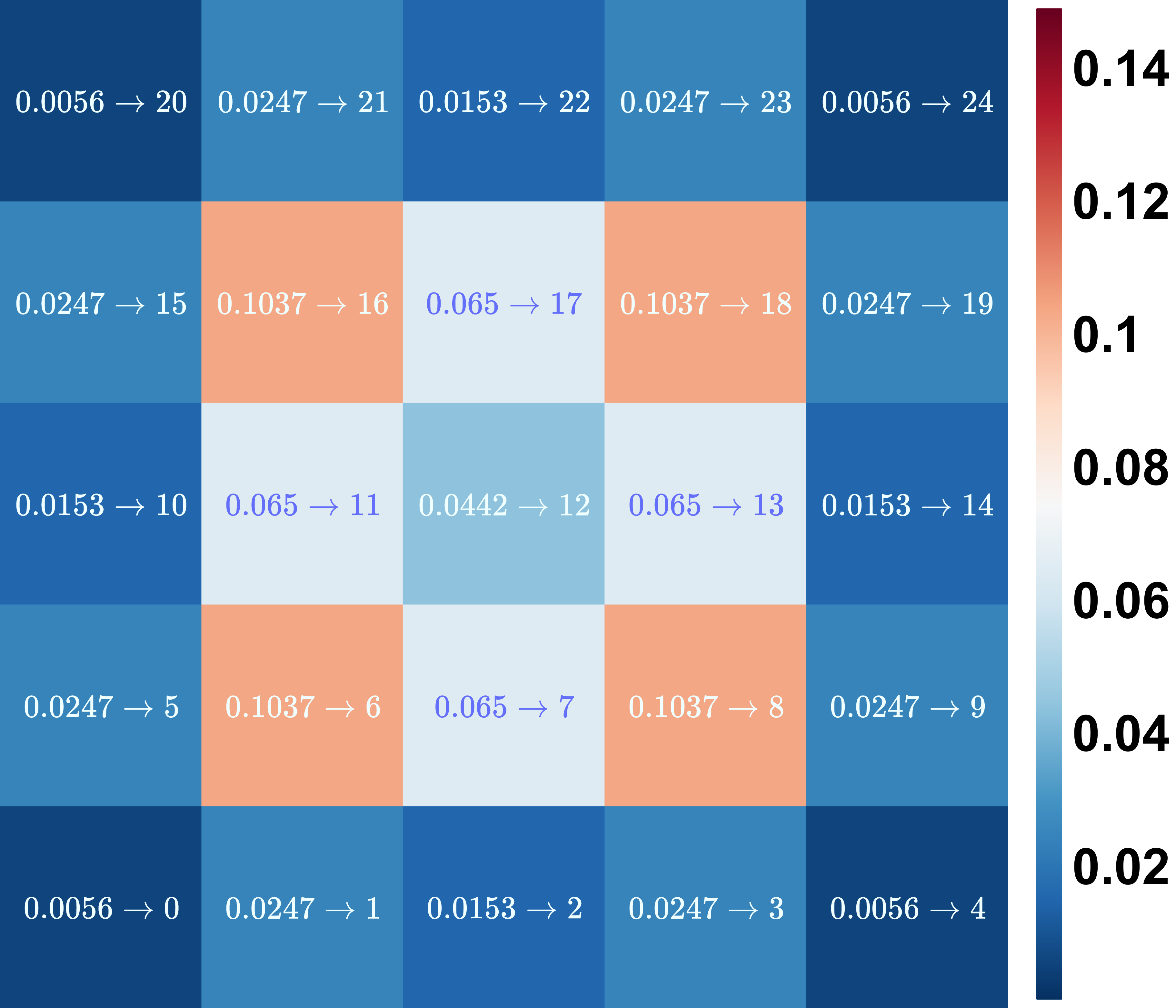}
  }
  \caption{The density distribution of the holon after released from the center (site-12) for time $\tau=0.8/t$ on a $5\times 5$ lattice with open boundary condition:  Lattice sites are labelled as 1,2,3, etc.. They are represented by a square.  The number on site ${\bf R}$ is the holon density $n({\bf R}, \tau)$. (a) and (b) are the results for the $tJ$ and  $\sigma tJ$ model respectively. (c) is the result for the a spinless fermion on empty lattice. The quantum state is calculated by keeping up to eighth order in $\tau t$.  }
\end{figure}

\section{Numerical results }

\noindent {\bf (IV.1) Holon density and spin-spin correlation : } 
We have performed numerical calculations for the density distribution and spin correlations of the holon
after it has traveled over a time interval $\tau$.
We consider a $5\times 5$ square lattice. The lattice sites are labelled as 0 to 24 as shown in Fig.\ref{proba_tj}. The center is at site-12. Initially, the hole is held fix at the center. All other sites are occupied by a fermion. 
We have computed the AF ground state $G(\bs{\mu}, {\bf 0})$ using exact diagonalization  \cite{lin_exact_1990,sandvik_computational_2010}. 
This result allows us to calculate the wavefunction $\Psi( \bs{\mu}, {\bf R}; \tau)$ as an expansion of $\tau t$ as shown in Eq.(\ref{expand}) and (\ref{Gamma-sum});  and hence the holon distribution at time $\tau$, $n({\bf R}, \tau)={\cal N}^{-1} \sum_{\bs{\mu}}|\Psi(\bs{\mu}, {\bf R}; \tau)|^2$, where ${\cal N}$ is the normalization constant.  

In Figure 2(a) and 2(b), we show the holon distribution $n({\bf R}, \tau)$ for the $tJ$-model and the $\sigma tJ$-model respectively, with the wavefunction $\Psi( \bs{\mu}, {\bf R}; \tau)$ calculated upto 8-th order of $\tau t$.  Figure 2(a) shows that the holon propagation in the $tJ$-model is very anisotropic, being strongest in the $x$ or $y$ direction. This is due to  the destructive interference of the strings  when the holon travel at an angle less then 90 degrees from the x (or y) axis as discussed in the previous section. 
In contrast, Figure 2(b) shows that the holon propagation in the $\sigma tJ$-model is much less anisotropic. In fact, it resembles more the density distribution  of a spinless fermion on an empty lattice, shown in Figure 2(c). 
This is because neither $\sigma tJ$-model nor the spinless fermion models has a Marshall phase in holon propagation, in contrast to the $tJ$-model. 

Figure 3(a) to 3(c) show the density $n({\bf R}, \tau)$ at different time $\tau t$ when the holon is found 
is at the center ${\bf 0}$, at the nearest neighbor and  at the next nearest neighbor from the center. 
When  $\tau t$ reaches 0.8, the difference between the $tJ$- and the $\sigma tJ$- model is very apparent when the final position  ${\bf R}$ is the nearest neighbor and the next nearest neighbor. See Figure 3(b) and 3(c). These differences should be experimentally measurable. Note also that the holon density of the $\sigma tJ$-model is closer to the spinless fermion than to the $tJ$-model.  The return probabilities to the origin of all three cases are quite similar, as shown in Figure 3(a). \\

\begin{figure}
  \subfigure[]{
    \includegraphics[width = 0.35\textwidth]{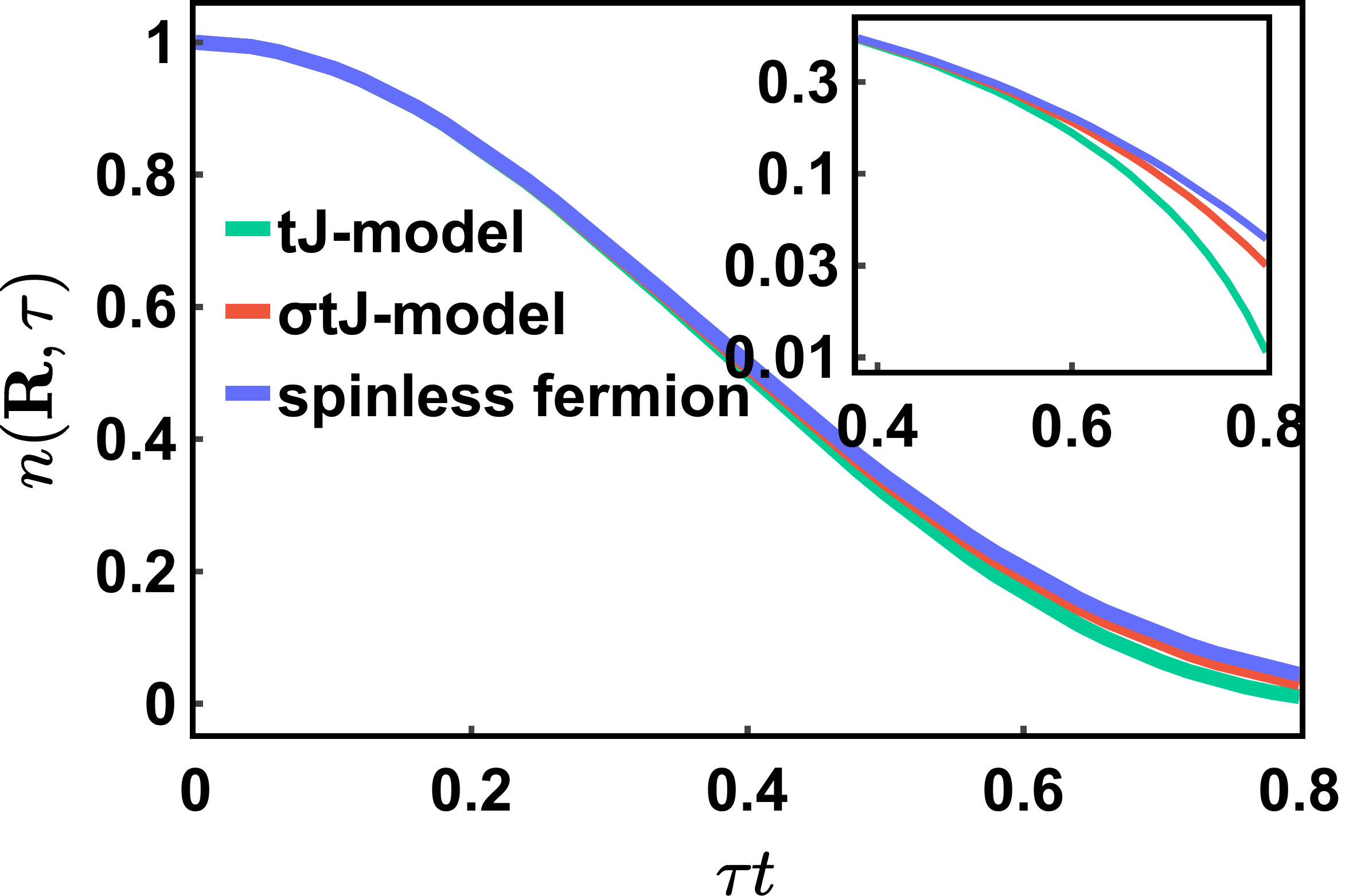}
  }
  \subfigure[]{
    \includegraphics[width = 0.35\textwidth]{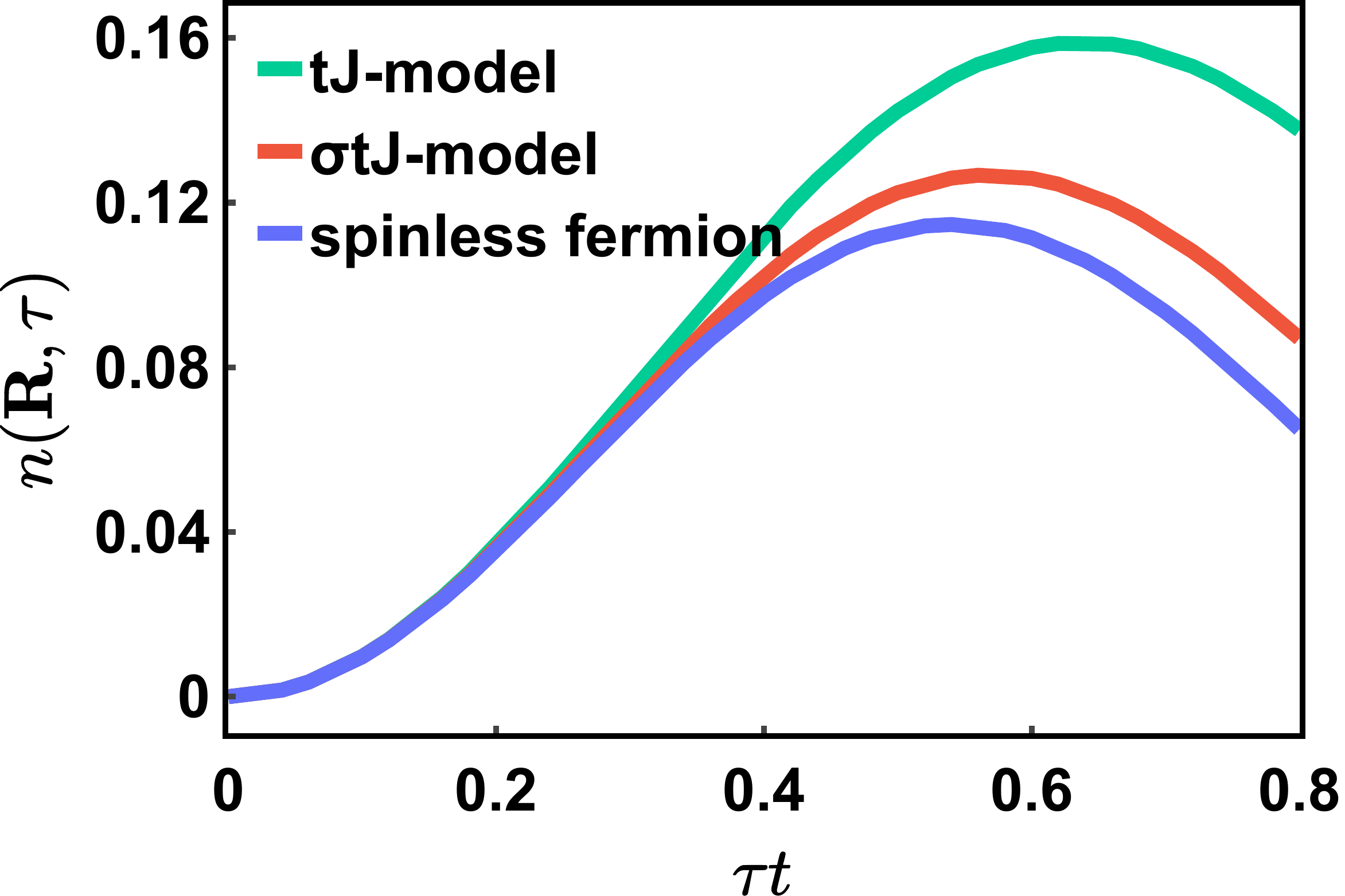}
  }
  \subfigure[]{
    \includegraphics[width = 0.35\textwidth]{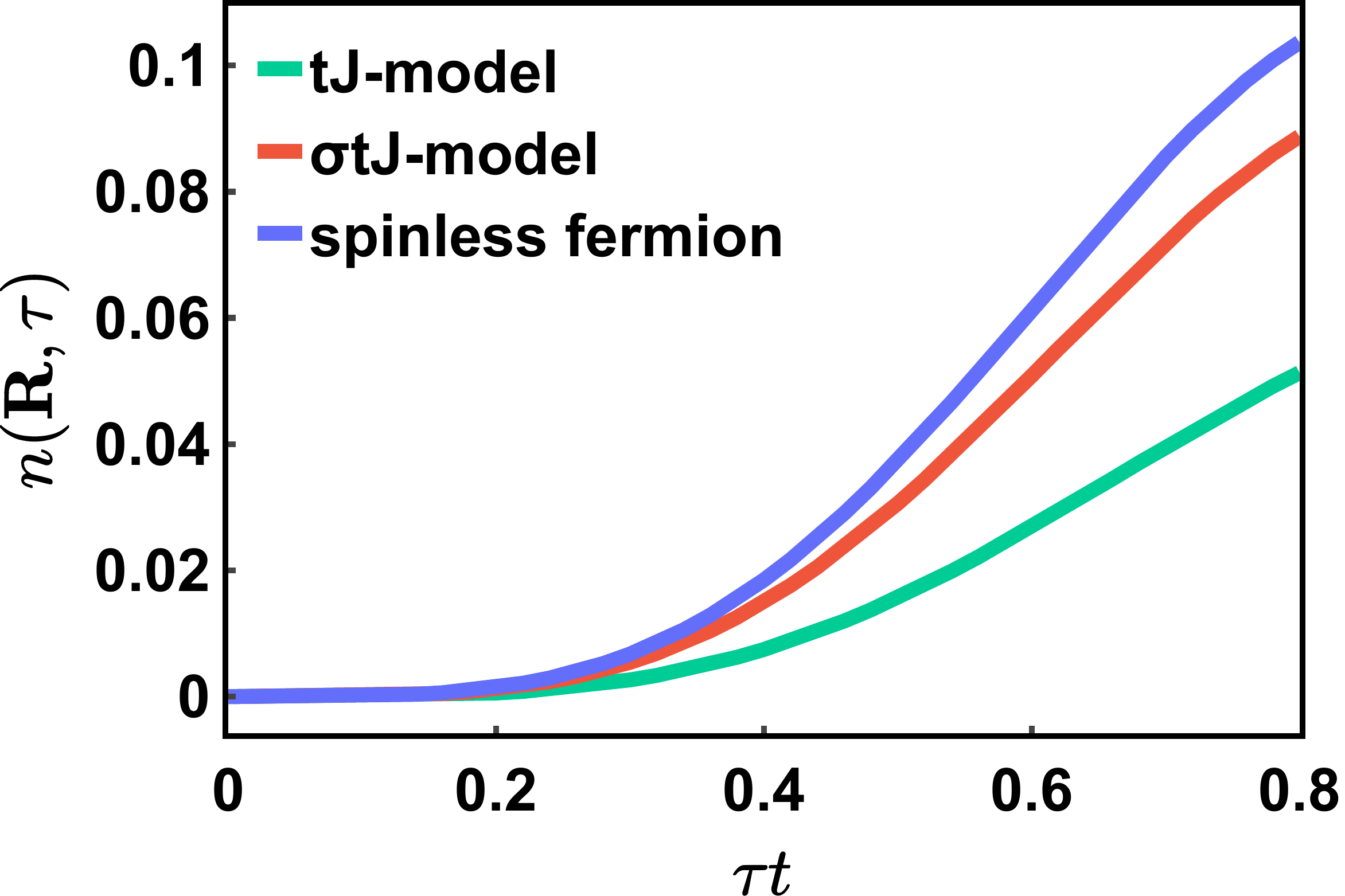}
  }
  \caption{The densities $n({\bf R}, \tau)$ in Figure 2 are plotted as a function of $\tau$ for different position ${\bf R}$.   (a) The holon returns to the origin (site-12).
  (b) The holon arrives at the nearest neighbor (site-13). (c) The holon arrives at the next nearest neighbor (site-18). 
   The results for holons in the $tJ$-model and the $\sigma tJ$-model are shown in green and red. 
   The result for a spinless fermion in an empty lattice is shown in blue. 
   At time $\tau=0.8/t$, the distinction between $tJ$-model and the $\sigma tJ$-model is apparent. 
  For the $tJ$-model, the holon propagates more efficiently along $x$ (case (b)) than along the diagonal (case (c)), with probability ratio $\sim 0.14/0.05$. In contrast, the corresponding ratio for the $\sigma tJ$-model is $\sim 0.08/0.08$. As explained in the text, the strong anisotropy of holon propagation in the $tJ$-model is due to the destructive interference of the holon strings. 
 It is also pointed out in the text that the propagation of a holon in the $\sigma tJ$-model (red curve) is similar to that of a single fermion in an empty lattice (blue curve) because the propagators in both cases have similar phase coherence. 
  In (a), the $t \tau \in [0.4, 0.8]$ part of the main plot is plotted in the inset using log scale.}
  \label{proba_t_tau}
\end{figure}

\begin{figure}
  \subfigure[]{\label{sz_sz_init}
    \includegraphics[width = 0.36\textwidth]{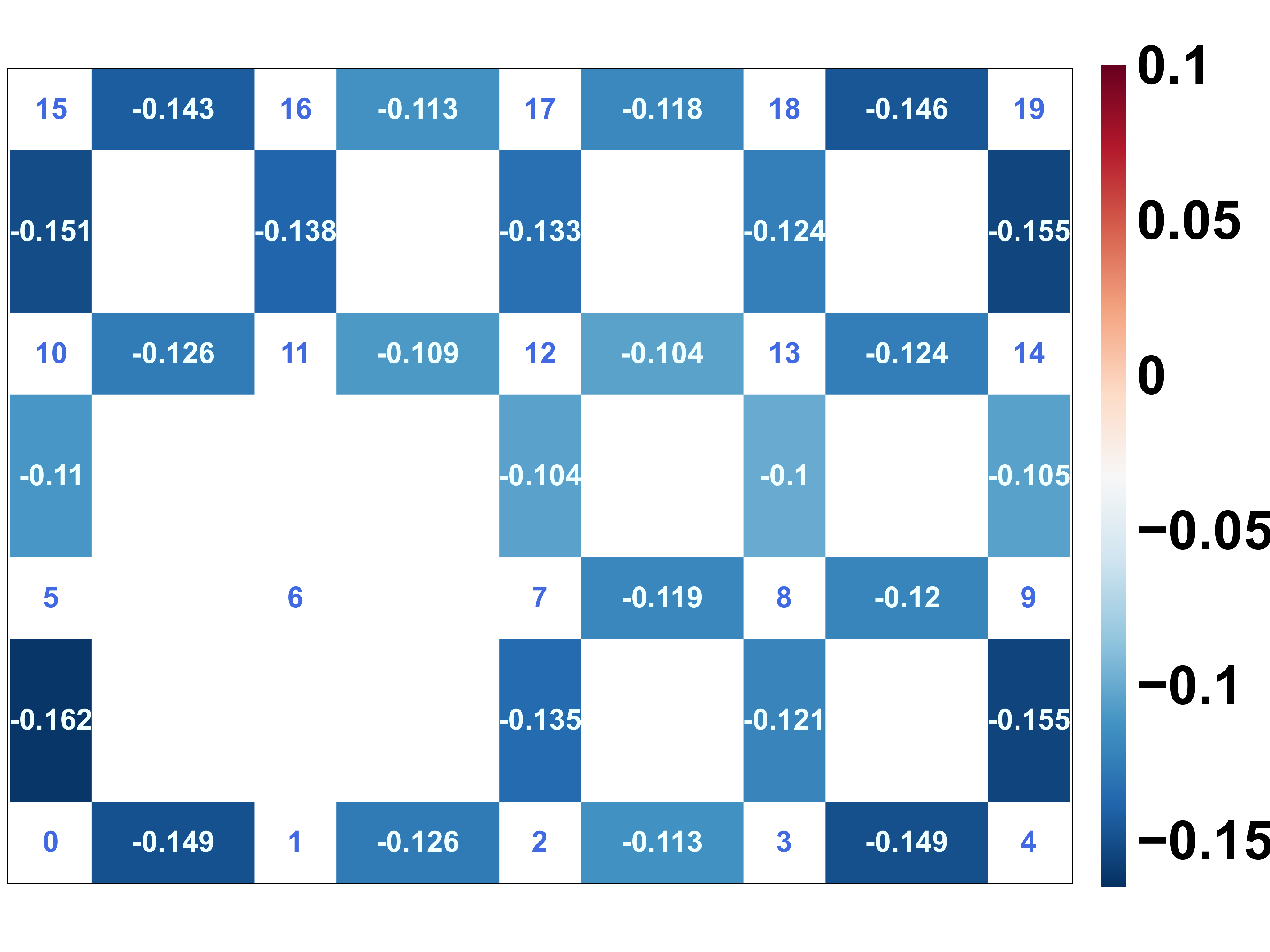}
  }
  
  \vspace{-0.4cm}
  
  \subfigure[]{\label{sz_sz_horiz}
    \includegraphics[width = 0.36\textwidth]{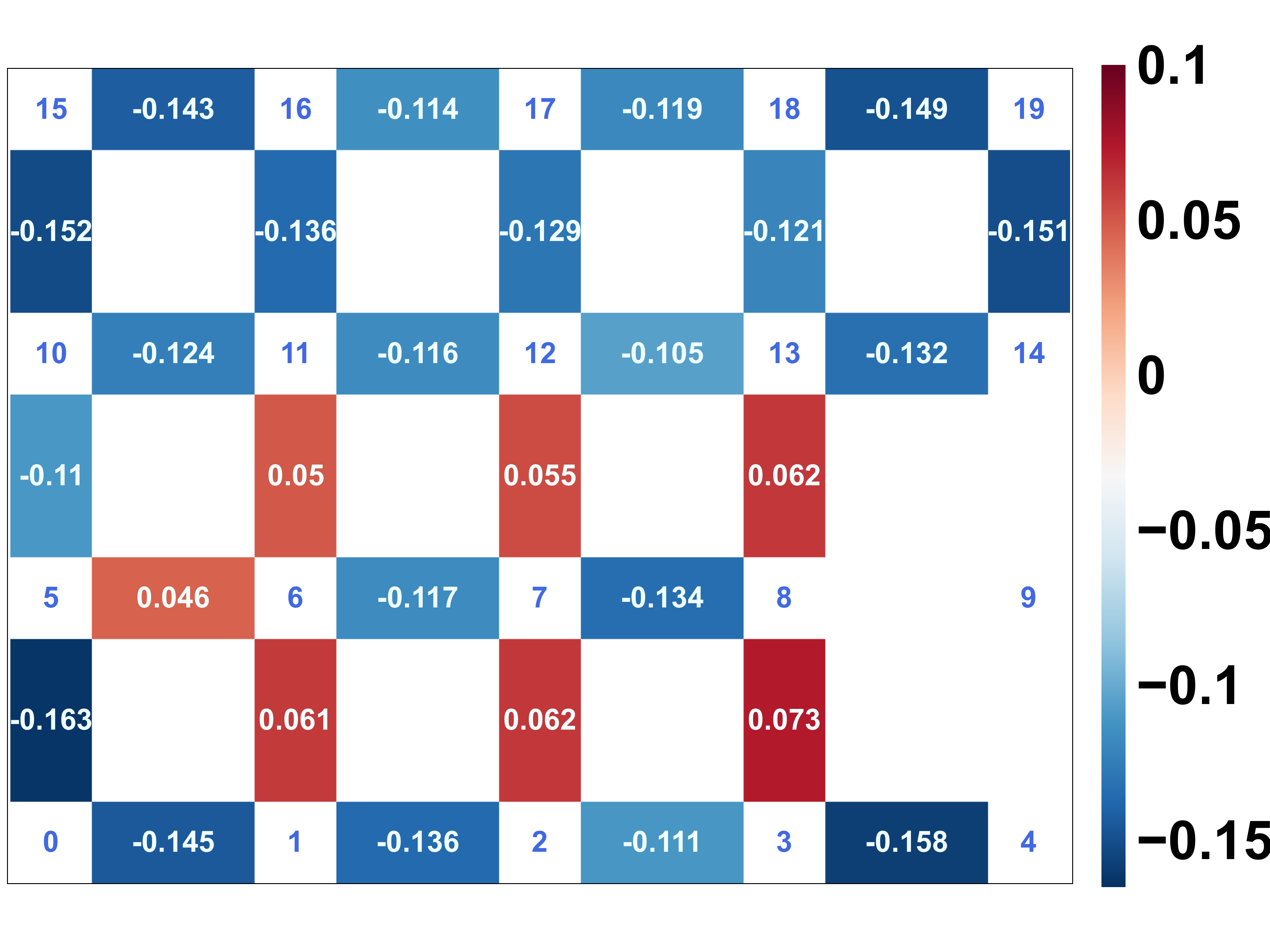}
  }
    
  \vspace{-0.4cm}
  
  \subfigure[]{
    \includegraphics[width = 0.36\textwidth]{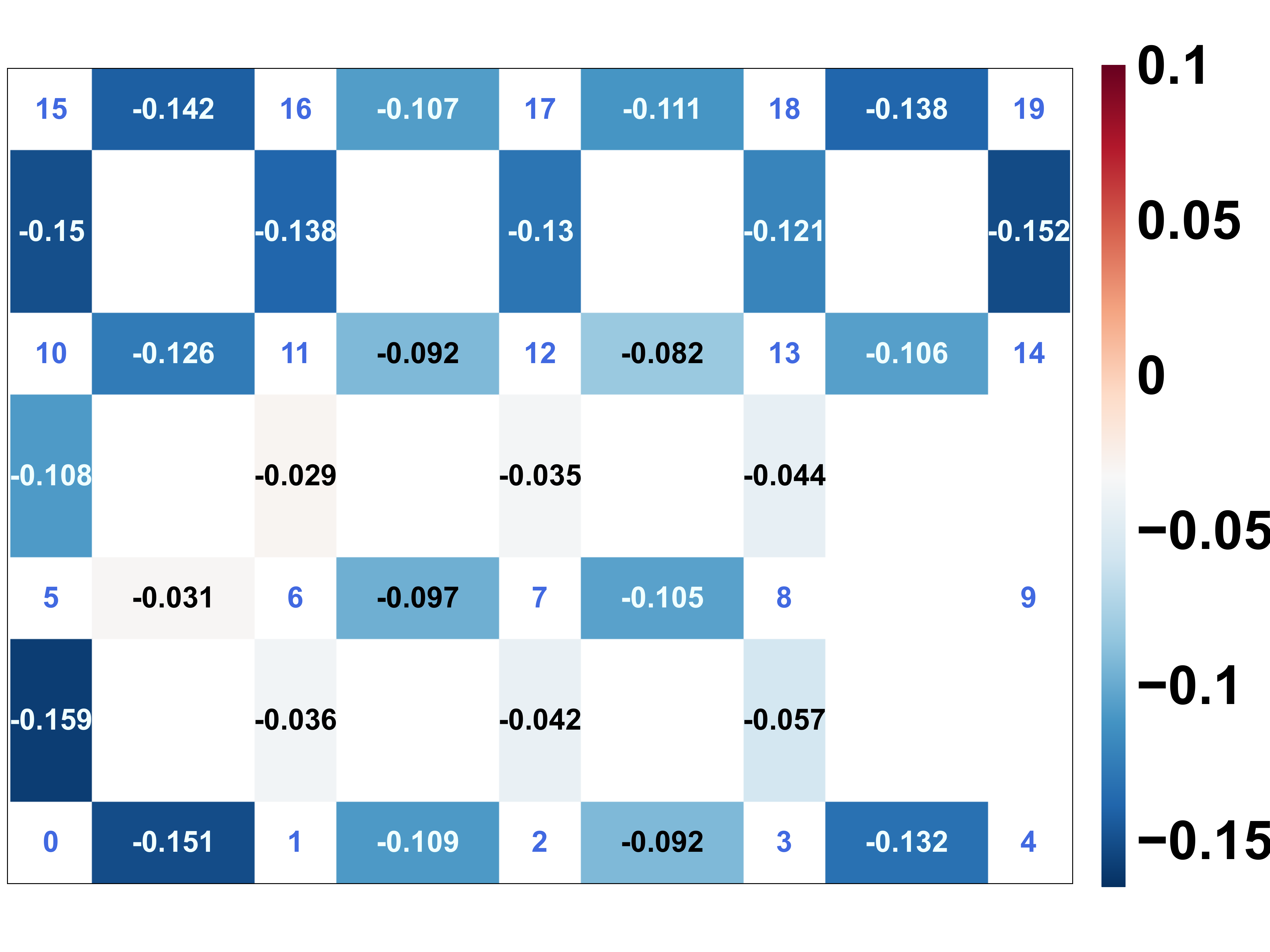}
  }
  \caption{Distribution of the exchange overlap $\rho_{ij}$ (or spin correlation)  when the holon is at different locations ${\bf R}$ at time $\tau=0.8/t$:  In this $5\times 4$ lattice, the small squares (labelled 0,1,2, etc) represent the lattice sites. The rectangle (or ``bond") between two neighboring squares ($i,j$) and the number on it represent the exchange overlap (or spin correlation ) $\rho_{ij}$. The quantum state after time $\tau$ is calculated by expanding Eq.(\ref{expand}) up to seventh order in $\tau t$.  (a) $\rho_{ij}$ of the initial state with the holon at site-6 (${\bf R}={\bf 0}$): Since the initial state is the AF ground state, it satisfies Marshall sign rule. Hence all the bonds ($\rho_{ij}$) are negative. (b)  $\rho_{ij}$ of the final state with the holon at site-9: Our numerical results have confirmed the picture discussed in Section I. All bonds not touching the string as well as all bonds on the string are antiferromagnet (i.e. $\rho_{ij}<0$), while the bonds at the immediate vicinity of the string is ferromagnetic  (i.e. $\rho_{ij}>0$). (c) Similar plot of (b) for the $\sigma tJ$-model: All the bonds remain negative as all the strings of the same length are phase coherent, as discussed in Section I.}
\end{figure}

\begin{figure}
  \subfigure[]{\label{sz_sz_final}
    \includegraphics[width = 0.36\textwidth]{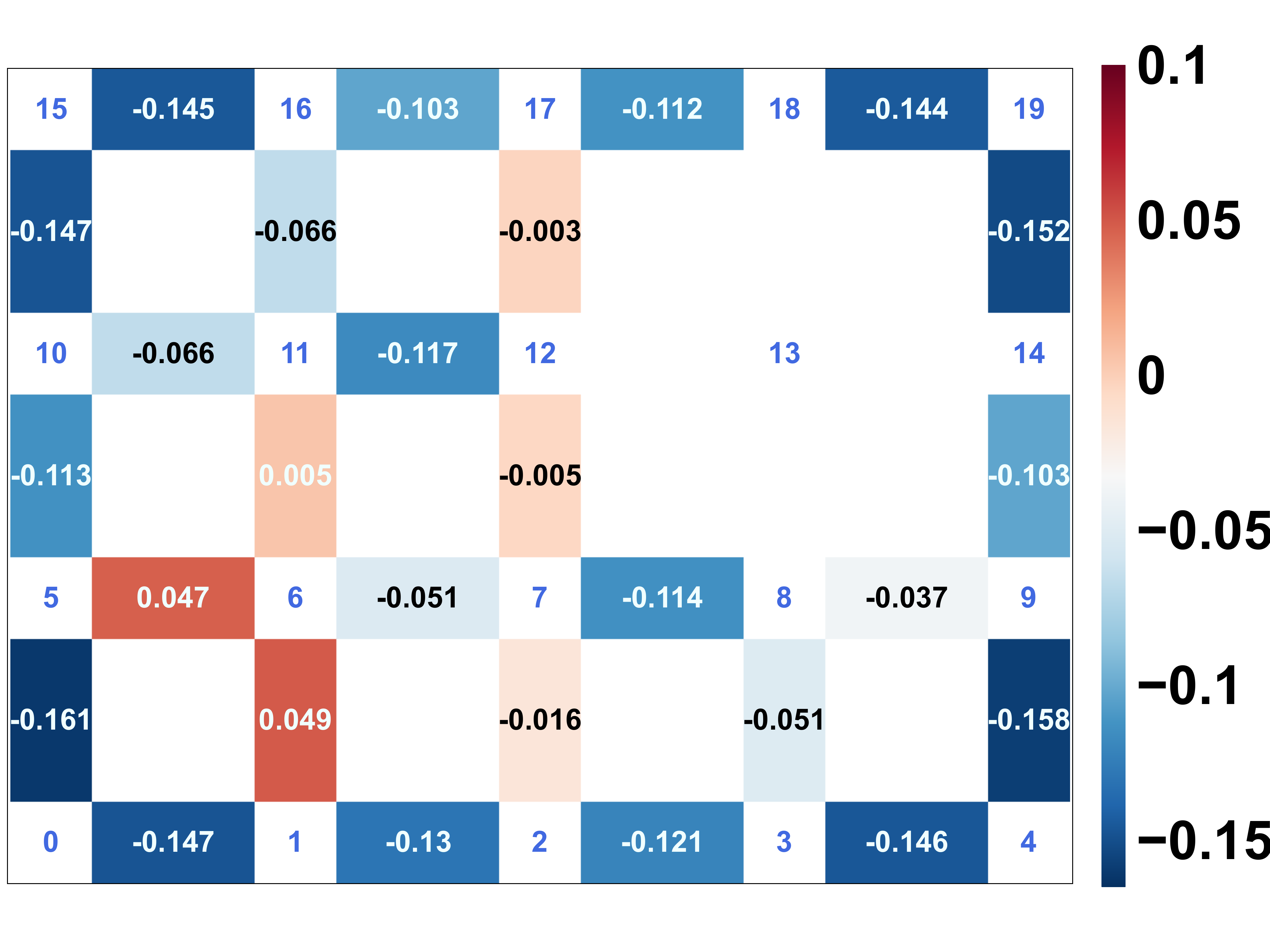}
  }
  \subfigure[]{\label{sz_sz_compare}
    \includegraphics[width = 0.36\textwidth]{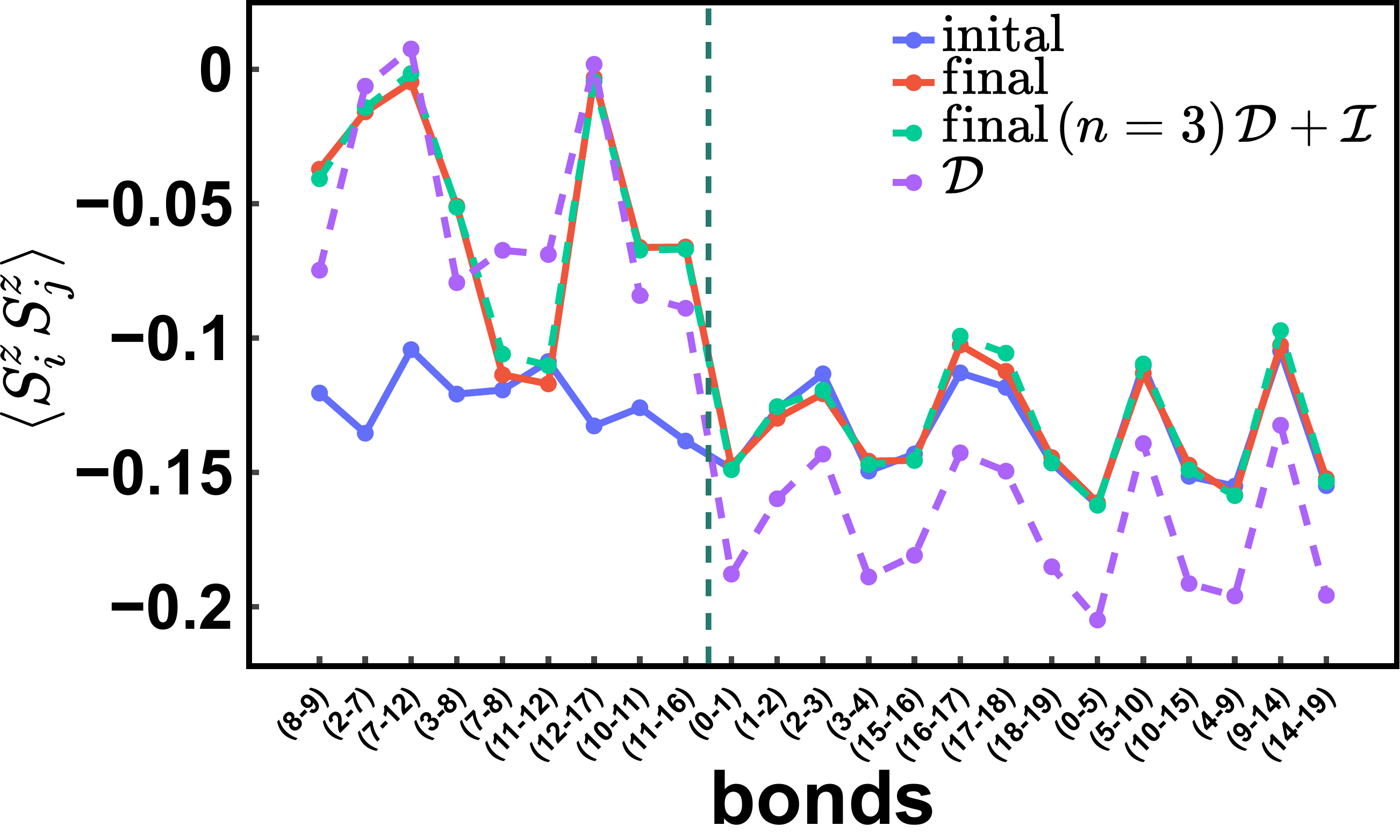}
  }
  \caption{(a) The distribution of spin correlation of the state at time $\tau = 0.8/t$ with the holon at site $n=13$, starting from the state at Fig.\ref{sz_sz_init}. The shortest string connecting the initial and final position (site-6 and site-13) has length $n=3$. These $n=3$ strings sweep through a rectangle $\Lambda$ covering sites 6, 7, 8, 11, 12, 13. For nearest neighbors $\langle i, j\rangle$ that 
  are inside $\Lambda$, such as (6,11), (7,12), as well as those connected to it, such as (7,2), (12,17), their spin correlations are strongly suppressed, due to the destructive interference of the holon strings. 
  (b) Comparison of spin correlations between the initial state at Fig.\ref{sz_sz_init}, and final state with final holon at site $n = 13$ at time $t \tau = 0.8$. The horizontal axis labels of neighboring pairs $\langle i, j\rangle$ of interest.  The final state is calculated by expanding Eq.(\ref{expand}) up to 7th order in $\tau t$. The result for the final state calculated with the only the shortest strings ($n=3$) in Eq.(\ref{expand}) is shown is green. Its closeness to the 7th order result (red dots) shows the dominance of the $n=3$ terms over the $n=5$ and 7 terms.  To the right and the left of the vertical dashed line are the neighboring pairs outside $\Lambda$ and those overlapping with it. For those to the right (outside $\Lambda$) , the spin correlations of the final state is essentially unchanged from the initial ones. For those to the left  (overlapping with $\Lambda$), the differences in spin correlation between the initial and final state are significant. This is the result of destructive interference of the strings that pass though one of the site (or both) in the neighboring pair $\langle i, j\rangle$ as discussed in Section II.  As discussed in Section III, the spin correlation of the final state (the green dots) is made up of ``diagonal"  and ``off-diagonal" terms, $\langle S_{i}^{z}S_{j}^{z}\rangle = {\cal D}_{ij} + {\cal I}_{ij}$.  Here, we have also plotted the direct term in purple. One  sees that ${\cal D}$ is close to the full value ${\cal D} + {\cal I}$ for the nearest neighbors that overlap with $\Lambda$ (i.e. those to the right of the vertical dashed line). It means the interference of holon string is close to complete  destruction, with ${\cal I}\sim 0$.}
\end{figure}

\noindent {\bf (IV.2) Marshall phase and spin correlations: } 
With the holon wavefunction $\Psi(\bs{\mu}, {\bf R};  \tau)$ calculated in {\bf (IV.1)}, we can evaluate the spin correlation $\langle S^z_i S^z_j \rangle$ for neighboring sites $\langle i, j\rangle$.  In Figure 4(a), we show this correlation of the initial AF ground state $|G; {\bf 0}\rangle$ with the immobile hole at site-6 (which is the origin ${\bf 0})$. The small squares represent the lattice sites. The number on the rectangle linking  two sites is the value spin correlation of these two sites. Blue and red color correspond to negative and positive sign. For this AF state, the correlations of all neighboring sites are negative, consistent with the Marshall sign rule.  The typical magnitude of the correlation is around -0.12. 

Figure 4(b) shows the spin correlation when the final position of the holon  is along $x$ (at site-9). In this case, the shortest path connecting the initial and final position is a straight line. 
The numerical result confirms the features discussed in the previous section, and in ref.\cite{ho_imaging_2020}, i.e.  $\langle S^z_i S^z_j \rangle$  is negative if   $i$ and $j$ are both outside or  on the string; and is positive if on one of them ($i$ or $j$) is on the string. The magnitude of the ferromagentic correlations (the red bonds) is weaker than that of the background AF by a fact of 2.  The exchange overlap of the $\sigma tJ$-model  (Figure 4(c)) is very different. It remains negative for all neighboring sites. 
This is due to the phase coherence of all strings of the same length as discussed before.  
The magnitude of the spin correlation in the vicinity of the string, however, is weaker than that of the AF background by roughly a factor of 4.

Figure \ref{sz_sz_final} shows the spin correlation when the final position of the holon (site-13) is not on the $x$-axis. In this case,  the strings connecting site-6 to site-13 have lengths $n=3, 5, 7, ..$. The pattern of spin correlation is very different from that in  Figure \ref{sz_sz_horiz} when the final position is along x. We note that the spin correlations of some neighboring sites (such as the ``bonds" between (17,12), (12,7), (7,2), (11, 6)) are reduced from the original AF value by almost a factor of 20; as if the system is driven towards a spin liquid state.   This large reduction shows the strong  destructive interference effects of the holon strings. 

In Figure \ref{sz_sz_compare}, we show the different contributions that make up the spin-spin correlation for the final state when the hole is at site-13 at time $\tau  = 0.8/t$. The horizontal axis shows the nearest neighbor pair of interest.  The values of the spin correlations of the initial state and the  final state (when the hole is at site-6 and site-13) are given in blue and red color respectively. These values are calculated from Eq.(\ref{Gamma-sum}) by including up to seventh order in $\tau t$.  On the same figure, we also show (in green color) the corresponding values for the final state including only the shortest strings (i.e. $n=n^{\ast}=3$) contributions. The fact that the blue and green dots almost overlap with each other shows that the $n=5$ and $n=7$-strings are less important. 

The $n=3$ strings connecting site-6 to site-13 sweep through a rectangle $\Lambda$ of size $2\times 1$ (including sites - 6, 7, 8, 11, 12, 13). 
The pairs of neighboring sites inside (and outside)
of $\Lambda$ are collected on the left (and the right) hand
side of the vertical dashed line.
 For the pairs outside $\Lambda$, there are little differences in the spin correlations between the initial  and the final states, (i.e. red and blue dots).   In contrast, the differences are significant for the neighboring sites inside $\Lambda$, a feature discussed in the previous section. 

To show quantitatively the different contributions to the string interference, we write the quantum state in Eq.(\ref{expand}) as $|\Psi^{(n)}\rangle= \sum_{\alpha}|\Psi^{(n)}_{\alpha}\rangle$, where ``$\alpha$" labels the strings.  The spin correlation can be written as $\langle \Psi^{(n)}|S_i^z S_j^z|\Psi^{(n)} \rangle =\cal{D} + \cal{I} $, where ${\cal D}$ and ${\cal I}$  are the ``diagonal" and the ``off-diagonal"  terms, 
${\cal D}= \sum_{\alpha} \langle \Psi^{(n)}_{\alpha}|S^{z}_{i} S^{z}_{j}|\Psi^{(n)}_{\alpha}\rangle$, and ${\cal I}=  
\sum_{\alpha \neq \beta} \sum_{\alpha} \langle \Psi^{(n)}_{\alpha}|S^{z}_{i} S^{z}_{j}|\Psi^{(n)}_{\beta}\rangle$. The off-diagonal terms represent the interference of the holon strings. 
In Figure \ref{sz_sz_compare}, the spin correlation ${\cal D} +{\cal I}$ and the diagonal terms are plotted as green and purple dots respectively. We see that these values are close to each other for the neighboring sites overlapping with $\Lambda$, i.e. those pairs on left hand side of vertical dashed line.  This means the interference contribution ${\cal I}$ is small, precisely the destructive interference physics we have discussed before when one or both of neighboring sites ($i, j$) are inside $\Lambda$.  
On the other hand, we have shown previously that the spin configurations of the final and  the initial state are identical outside $\Lambda$, independent of the strings inside it. This means both diagonal  and off-diagonal  contributions are present, and their sum is equal the original AF correlation, as seen on agreement between the green and red data in  Figure \ref{sz_sz_compare}.

\section{Conclusion}
We have studied the motion of the a single hole in the $tJ$-model in the limit of slow spin motion, $t/J\ll 1$. In this limit, the propagation of a holon generates  strings equipped with a Marshall phase, which depends on the spin configurations in the underlying antiferromagnetic state.  The interference of these strings leads to a holon propagation much more anisotropic than the that of free fermions. It also 
reduces substantially the antiferromagnetic correlation 
in the region swept through by the strings.  We further demonstrate the effect of the Marshall phase by considering the so-called $\sigma tJ$-model.  The spin dependent hopping of this model  removes completely the Marshall phase of the antiferromagnet, making the holon propagation similar to that of a spinless fermion. Our method can be generalized to study  multi-holon transport, and include the effect of spinons, which we shall discuss elsewhere.   \\

Acknowledgement: We thank Professor Zheng-Yu Weng for helpful discussions, and the kind support of the MacMaster Funds. 

\bibliography{holon}

\end{document}